\documentstyle[emulateapj,rotate,psfig]{article}

\def\eg{{\it e.g.}}
\def\etal{{\it et al.}}
\def\etc{{\it etc.}}
\def\ie{{\it i.e.}}
\def\fbar{\overline{f}}
\def\fm{f_{\rm max}}
\def\ffg{f_{\rm fine}}
\def\fcg{f_{\rm coarse}}
\def\spose#1{\hbox to 0pt{#1\hss}}
\def\ltsim{\mathrel{\spose{\lower.5ex\hbox{$\mathchar"218$}}
     \raise.4ex\hbox{$\mathchar"13C$}}}
\def\gtsim{\mathrel{\spose{\lower.5ex \hbox{$\mathchar"218$}}
     \raise.4ex\hbox{$\mathchar"13E$}}}

\slugcomment{Rutgers Astrophysics Preprint \#271 \\ Revised version to appear in 
\it The Astrophysical Journal Letters}

\begin{document}

\title{Is a Simple Collisionless Relic Dark Matter Particle Ruled Out?}
\author{J. A. Sellwood}
\affil{Department of Physics and Astronomy, Rutgers University, \\ 
136 Frelinghuysen Road, Piscataway NJ 08854 \\
sellwood@physics.rutgers.edu}

\begin{abstract}
The central densities of dark matter (DM) halos are much lower than predicted in 
cold DM models of structure formation.  Confirmation that they have cores with a 
finite central density would allow us to rule out many popular types of 
collisionless particle as candidates for DM.  Any model that leads to cusped 
halos (such as cold DM) is already facing serious difficulties on small scales 
and hot DM models have been excluded.  Here I show that fermionic warm DM is 
inconsistent with the wide range of phase space densities in the DM halos of 
well-observed nearby galaxies.


\end{abstract}

\section{Introduction}
The well-known mass discrepancies in galaxies (Bosma 1978; Broeils 1992; 
Verheijen 1997) and in clusters of galaxies (Zwicky 1937; Carlberg, Yee \& 
Ellingson 1997; Tyson, Kochanski \& Dell'Antonio 1998) are usually taken to 
imply the existence of a large fraction of invisible, or ``dark,'' matter (DM) 
in the universe.  A popular candidate is Cold Dark Matter, for which the DM 
particles, whatever they are, are almost at rest with respect to the Hubble flow 
in the early universe.  CDM is often imagined to be a heavy, non-baryonic relic 
particle from the early universe which has essentially only gravitational 
interactions with itself and with normal, or baryonic, matter.

The CDM model has been studied intensively for twenty years and now has many 
well-worked out predictions for the formation of structure in the universe (\eg\ 
Bertschinger 1998).  The broad-brush impression one now has is that the 
currently favored $\Lambda$CDM model boasts a considerable degree of success in 
predicting large-scale structure (\eg\ Pearce \etal\ 1999; Bahcall \etal\ 1999). 
 But it has become apparent in recent years that the predictions of almost any 
flavor of CDM are seriously at variance with the observed properties of galaxies 
because the central densities of collapsed objects and fragments are predicted 
to be too high (\S\ref{sec:cusps}).

A number of authors have therefore begun to explore variations of the CDM model; 
most favor a modification to the properties of the DM particles rather than the 
alternative -- a change to the law of gravity.  The simplest is the warm DM 
matter model (\eg\ Colombi, Dodelson \& Widrow 1996; Sommer-Larsen \& Dolgov 
1999; Hogan 1999) in which streaming of the DM particles in the early universe 
suppresses small-scale power in the fluctuation spectrum.  In addition, WDM 
particles in halos of greater volume density must have larger velocity spreads, 
because of Liouville's theorem, thereby precluding strong density gradients.

Simulators of the WDM model suppress small-scale power in the fluctuation 
spectrum but generally ignore the initial finite velocity spread, which is 
difficult to include without wrecking the quiet start.  Since the DM in these 
simulations still has infinite phase space density, the resulting halo profiles 
have density cusps resembling those which form in CDM (\eg\ Moore \etal\ 1999; 
Col\'\i n \etal\ 2000).  

\section{Phase space density constraint}
\subsection{Prediction}
If the DM particle decoupled from thermal equilibrium at some early epoch, then 
its phase space density today can be predicted.  Fermionic DM would have a 
finite initial maximum $f = \fm$, which leads the well-known constraint on 
neutrino masses (Tremaine \& Gunn 1979).

Hogan (1999) suggests that halos today may reflect, $\fbar$, the average over 
the initial velocity distribution.  But $\fm$ is unbounded for bosons, making a 
single value of $\fbar$ rather poorly defined (Madsen 2000).  Thus phase space 
density constraints are much weaker for bosonic WDM; careful simulations are 
needed to predict the structure of halos in this model, but the small fraction 
of high $f$ material makes mild cusps seem likely.

The fine-grained phase space density, $\ffg$, of a truly collisionless fluid is 
strictly conserved, but the maximum coarse-grain density, $\fcg$, can decrease 
during violent evolution, such as a collapse or merger.  Simulations, which by 
their nature can track only $\fcg$, have found, however, that the maximum $\fcg$ 
{\it barely changes\/} in even quite violent collapses (May \& van Albada 1984) 
or mergers (Farouki, Shapiro \& Duncan 1983; Barnes 1992).  These tests were 
with systems that have finite $\fm$ at the outset, whereas $\fcg$ can, and 
generally does, decrease during the collapse or merger of systems with unbounded 
$\fm$.  Hernquist, Spergel \& Heyl (1993) merged systems with cusped density 
profiles having unbounded $\fm$; they defined an ``average'' phase space density 
over the inner half of the mass distribution $\overline{f}_{1/2}$, which 
decreased during major mergers.  Dalcanton \& Hogan (2000) construct a heuristic 
argument that $\fcg$ should decrease during mergers, but like Hernquist \etal, 
their argument applies to the average value over the inner part of the halo, 
which probably does decrease, rather than the specific central value which, if 
originally finite, remains almost unchanged.

Processes such as baryonic cooling and smooth infall, or even feedback from star 
formation, do not change $f$ for the DM.  Scattering of halo particles by dense 
clumps of baryonic matter could change $f$ locally, but simulations with a 
baryonic component (\eg\ Navarro \& Steinmetz 2000) do not appear to lead to 
significantly different halo profiles from those in which baryonic processes are 
omitted.  Dynamical friction in barred galaxies also hardly affects the 
structure of the inner halo, even when it is intolerably fierce (Debattista \& 
Sellwood 2000).

Thus $\fm$ for halos today should all cluster around a common value if the DM 
particle is a collisionless fermion having some fixed velocity spread in the 
early universe.

\subsection{Phase Space Densities of DM Halos}
\label{sec:psd}
Fermionic WDM halos should today approximate cored isothermal spheres having 
finite central densities.  The 1-D velocity dispersion in an isotropic, 
isothermal sphere is simply $\sigma = 2^{-1/2} V_{\rm flat}$, where $V_{\rm 
flat}$ is the flat circular speed from the halo (Binney \& Tremaine 1987 
\S4.4b).  A finite central density, $\rho_0$, yields a rotation curve with a 
central slope $dV/dR = (4\pi G\rho_0/3)^{1/2}$.  For the popular halo fitting 
function $\rho = \rho_0 / (1 + r^2 / r_0^2)$, the asymptotic velocity, core 
radius and central density are related as $V_{\rm flat} = (4\pi G\rho_0)^{1/2} 
r_0$.  Other functional forms (\eg\ Evans 1993), departures from sphericity, or 
from velocity isotropy merely introduce correction factors of order unity.  Thus 
measurements of $V_{\rm flat}$ and of either $\rho_0$ or $r_0$ for the halo 
allow us to estimate $\fm \sim \rho_0/\sigma^3$.

Rotation curves of galaxies include contributions from components other than DM, 
of course.  HI data that extend well outside the optical part of the galaxy 
provide reasonably firm values for $V_{\rm flat}$.  Rotation curves that are 
well-resolved in the inner parts provide an upper limit to $\rho_0$, which may 
be close to the actual value if the baryonic contribution is small, as in low 
surface brightness systems (LSBs).  In many cases, however, either $V_{\rm 
flat}$ or $\rho_0$ for the DM halo depends on the decomposition of the rotation 
curve into the separate contributions from luminous and dark matter, which is 
generally controversial.  I assume maximum disk models, and discuss below how 
this assumption affects the values.

\begin{figure}[p]
\centerline{\psfig{file=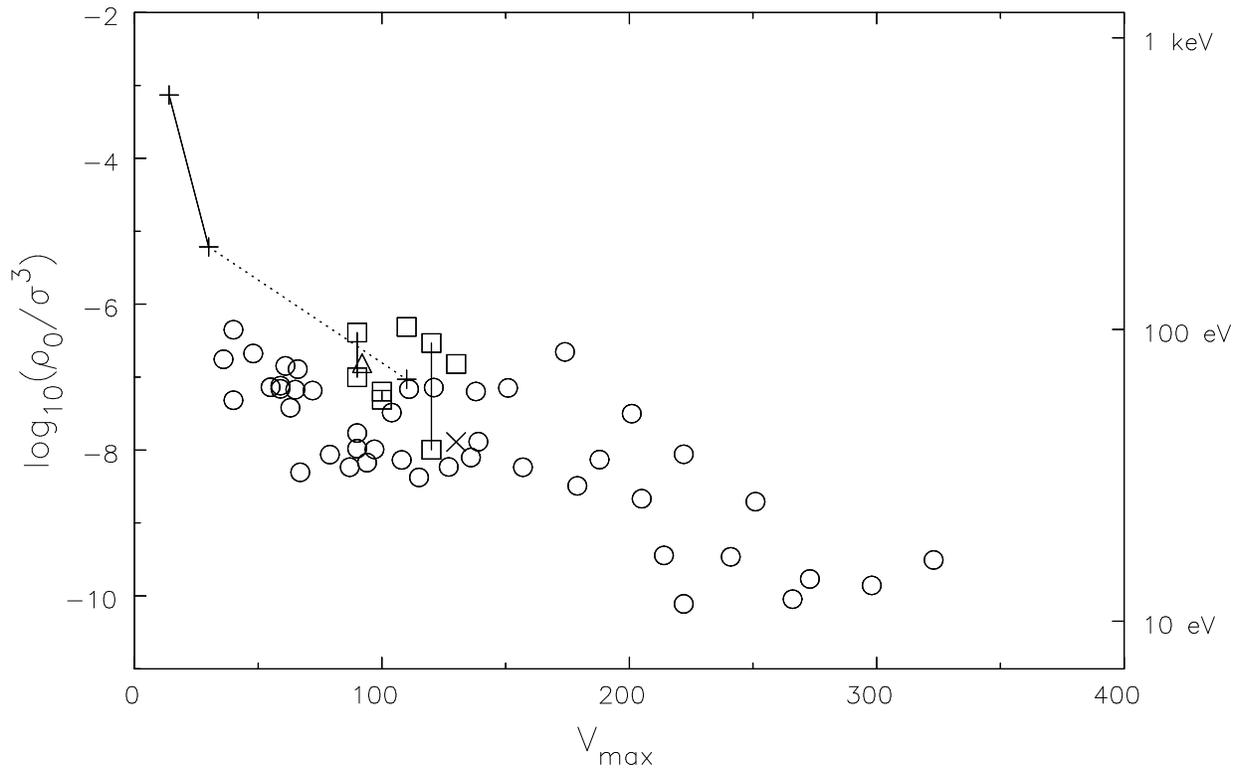,width=0.9\hsize,angle=0}}
\caption{Estimates of $\fm \;(\sim \rho_0/\sigma^3)$ for DM in a number of 
galaxies plotted against the observed $V_{\rm max}$.  The pluses and squares are 
described in the text, the circles are from Broeils (1992) and Verheijen (1997), 
the cross is for NGC 4123, and the triangle is for NGC 5585.  Table 
\ref{tab:data} gives the raw data and its source for each galaxy.  The units of 
$\fm$ are M$_\odot$ pc$^{-3}$ / (km s$^{-1}$)$^{-3}$ and km s$^{-1}$ for $V_{\rm 
max}$; the right-hand axis shows values of the rest mass of an equivalent 
thermal relic fermion.}
\label{fig:psd}
\end{figure}

Figure \ref{fig:psd} shows estimates of $\fm \sim \rho_0/\sigma^3$ in a number 
of well-observed galaxies.  The data and their sources are summarized in Table 
\ref{tab:data} with the exception of the points for the Draco and Ursa Minor 
dwarf spheroidal galaxies.  Broeils (1992) gives large formal values of $V_{\rm 
flat} \gg V_{\rm max}$ for three galaxies (noted in Table \ref{tab:data}); in 
these cases, I conservatively determine $\sigma$ from $V_{\rm max}$ instead of 
$V_{\rm flat}$.

The pluses in Figure \ref{fig:psd} are for the Draco and Ursa Minor dwarf 
spheroidal galaxies, which both have a stellar velocity dispersion $\sim 10$ km 
s$^{-1}$ (Armandroff, Olszewski \& Pryor 1995).  The uppermost point assumes a 
simple mass-follows-light King model (Binney \& Tremaine 1987, \S4.4) with an 
estimated King radius for Draco of 150~pc (Pryor \& Kormendy 1990) that yields a 
central density of 0.7~M$_\odot$ pc$^{-3}$ (Pryor, private communication).  The 
DM halo could have a larger $r_0$ and $\sigma$ than the stars, however, implying 
a lower $\fm$.  We place an extreme lower bound on $\fm$, which is essentially 
the same for both Draco and UMin., as follows: We adopt the lower bound $\rho_0 
= 0.2$ M$_\odot$ pc$^{-3}$ (Olszewski 1998), and the argument (Gerhard \& 
Spergel 1992) that their masses must be $< 10^{10}$ M$_\odot$.  Treating this 
gigantic, low-density halo as a $W_0=1$ King model, we obtain $\sigma_{\rm 
DM}\sim 130$ km s$^{-1}$.  Slightly more extreme models could be imagined, but 
M/L$_V \sim 30\,000$ for this model already!  A more resonable lower bound might 
be to adopt M/L$_V \sim 500$, on the high side for galaxy clusters (\eg\ 
Carlberg \etal\ 1997; Tyson \etal\ 1998), which would require $\sigma_{\rm 
DM}\sim 32$ km s$^{-1}$.  The values of $\fm$ for this, and the above more 
extreme, model are shown by the lower pluses in Figure \ref{fig:psd}.

While maximum disk models are not universally accepted, most of the points in 
Figure \ref{fig:psd} would not move much if this assumption were dropped.  Three 
of the LSBs from Swaters, Madore \& Trewhella (2000) are plotted (squares) for 
both their ``maximum disk'' and ``no disk'' fits for which $\fm$ differs by at 
most factor 100.  The DM halo for a ``no disk'' fit to NGC 3198, the 
proto-typical galaxy for DM studies, has the same $V_{\rm flat}$ but a core some 
5-6 times smaller (van Albada \etal\ 1985), increasing $\fm$ by the square of 
this factor.  While these are substantial uncertainties, they are small in 
relation to the total range.

\subsection{Implication}
There are three possible conclusions from this Figure.  First, the range of 
$\fm$ could be mostly due to errors.  Clearly some invidual points could be in 
error by a couple of orders of magnitude, but others are more certain.  The 
spread is admittedly greatly increased by the highest point for the dwarf 
spheroidals, which is not well determined because we observe only the stars at 
the centers of the halos.  This point can be brought into the main cluster in 
the Figure, but only by adopting a truly extreme model.  Similarly, the lower 
points could be moved up by adopting sub-maximal disk models.  If one wishes to 
argue that the {\it entire\/} spread is due to measurement errors, and that the 
DM halos manifest a characteristic value for $\fm$, that value corresponds to a 
thermal relic fermion with a mass of $\ltsim 100\;$eV, \ie\ hot DM (HDM), which 
has been rejected previously (\eg\ Gerhard \& Spergel 1992; Cen \& Ostriker 
1992).

Dalcanton \& Hogan (2000), who present a similar diagram, argue that the 
measurements are not of $\fm$ but of some average value over the inner halo.  If 
the density profiles of all halos rise to a finite central value with a similar 
functional form, however, their quantity would also have a characteristic value 
in WDM.  To be consistent with WDM, they require the halo density to continue to 
rise steeply inside $r_0$, with high $f$ material at the very center 
contributing little to the average over the volume inside $r_0$.  Their 
interpretation therefore therefore applies to a DM particle that has initially 
unbounded $\fm$.

If the values in Figure 1 are indeed measurements of $\fm$, most of the seven 
orders of magnitude spread is enormously larger than could be allowed if the WDM 
particle were a collisionless fermion.

\section{Halos with cusps}
\label{sec:cusps}
Phase space density constraints cannot be applied to CDM, or strictly to bosonic 
WDM despite its velocity dispersion, since $\fm$ is unbounded.  It is the 
absence of an upper bound to the phase space density in CDM which gives rise to 
density cusps in the collapsed halos (Moore \etal\ 1998; Klypin \etal\ 2000).  
(This is also true of the cold component of the less-appealing mixed DM models 
[\eg\ Kofman \etal\ 1996].)  An unbounded $\fm$ is the root cause of the 
difficulties now besetting CDM:

(1) The ``concentration index'' of the halo, which has a range of values 
(Bullock \etal\ 1999), implies a high central DM density, that should increase 
still further as the baryons cool and settle to the center.  Low luminosity 
galaxies and LSBs are believed to have the largest fractions of DM, and 
therefore halos for which compression by baryonic infall is least important.  
Yet the rotation curves of these galaxies (C\^ot\'e \etal\ 1997; Swaters \etal\ 
2000) rise gently, suggesting a low density core, irrespective of the M/L 
ascribed to the baryonic component.  The halo profiles in bright HSB galaxies 
are more model-dependent, since the rotation curves in these galaxies generally 
do rise quickly (Rubin, Kenney \& Young 1997; Sofue \etal\ 1999).  It is now 
clear, however, that the inner part of the rotation curves of these galaxies is 
dominated by the luminous disk and bulge; even in these cases, the DM halo has a 
large, low-density core (Debattista \& Sellwood 1998) or very low concentration 
index (Weiner \etal\ 2000).  Thus, there is little support for high central 
densities of DM within the halos of any type of galaxy.

(2) The merging hierarchy causes the cooled baryonic fraction to lose angular 
momentum to the halo, making disks that are too small (Navarro \& White 1994; 
Navarro \& Steinmetz 1997).  The predicted angular momentum of the disk is an 
order of magnitude less than that observed.  The problem is only partially 
ameliorated (MacLow \& Ferrara 1999; Navarro \& Steinmetz 2000) if some process 
(usually described as ``feedback from star formation'') prevents most of the gas 
from cooling until after the galaxy is assembled.

(3) Navarro \& Steinmetz (2000) describe their failure to predict the zero-point 
of the Tully-Fisher relation as a ``fatal problem for the $\Lambda$CDM 
paradigm.''  They show that no matter what M/L is assumed for the disk, the 
predicted circular speed at a given luminosity is too high because the halo 
density is too high.  The Tully-Fisher prediction may be even worse, since CDM 
predicts $L \propto V^3$ (Dalcanton, Spergel \& Summers 1997; Mo, Mao \& White 
1998), whereas Verheijen (1997) stresses that when $V$ is interpreted as the 
circular velocity of the flat part of the rotation curve, the true relation is 
closer to $L \propto V^4$.

(4) The results from high-resolution $N$-body simulations (Klypin \etal\ 1999; 
Moore \etal\ 1999) have revealed large numbers of sub-clumps within large DM 
halos, many more than are observed as satellite galaxies.  Whether these 
fragments threaten the survival of thin disks in the host galaxy, remains to be 
seen.  (WDM models are largely motivated to avoid this problem by suppressing 
small-scale power in the initial perturbation spectrum.)

These difficulties may not yet be fatal to CDM, since a better understanding of 
baryonic processes in galaxy formation could conceivably alter the predictions.  
It is unclear what process could loosen the high density clumps of DM, however; 
\eg\ Debattista \& Sellwood (2000) show that an intolerable degree of dynamical 
friction has a very mild effect on the halo density profile.  

The first of these problems, which is shared by all DM models that predict 
cusped halos, is key.  The highest quality data (\eg\ Blais-Ouellette \etal\ 
1999) do appear to show that the halo density profile rises gently to a finite 
central value.  This result is controversial, however, since others (\eg\ van 
den Bosch \etal\ 1999) argue that the observed rotation curves could be 
consistent with mild density cusps in the halos.  More high-quality data should 
eventually settle the issue.

\section{Conclusions}
\label{sec:concl}
Fermionic WDM halos should have well-defined cores with a characteristic $\fm$. 
This prediction is inconsistent with the range of values shown in Figure 1 which 
rules out this particular DM candidate.  Halos having finite central densities 
are also inconsistent with the density cusps predicted by CDM, and possibly also 
bosonic WDM.  Other work has excluded HDM.  Thus the existence of finite central 
density cores in halos would rule out most forms of simple, collisionless relic 
particle as DM candidates.  It is therefore of great importance to obtain 
tighter observational constraints on the inner density profiles of halos.

It is suddenly popular to hypothesize extra properties for the CDM particle in 
order to soften the collapsed halos (Spergel \& Steinhardt 1999; Peebles 2000; 
Goodman 2000; Hu \etal\ 2000; Kaplinghat \etal\ 2000; \etc).  The simple CDM 
model would become much less attractive if an extra {\it ad hoc\/} property were 
needed to rescue it.

The nature of DM is increasingly constrained by the observed properties of 
galaxy halos.  In particular, any successful theory of galaxy formation will 
need to account for the wide range of phase space densities for DM shown in 
Figure 1.

\acknowledgments  I would like to thank Tad Pryor, Arthur Kosowsky, Terry 
Matilsky and Julianne Dalcanton for lively discussions and Scott Tremaine, Josh 
Barnes and Stacy McGaugh for helpful e-mails.  This work was supported by NSF 
grant AST 96/17088 and NASA LTSA grant NAG 5-6037.

\begin{deluxetable}{lrrrrc}
\tablewidth{0pt}
\tablecaption{Galaxies Plotted in Figure \ref{fig:psd}}
\scriptsize
\tablehead{
\colhead{Name} & 
\colhead{$r_o$\tablenotemark{a}} & 
\colhead{$V_{\rm flat}$\tablenotemark{b}}  & 
\colhead{$\rho$\tablenotemark{c}} & 
\colhead{$V_{\rm max}$\tablenotemark{d}}  & 
\colhead{Ref} } 

\startdata
\nl
DDO  154 &   2.0 &   59 &  15.3 &  48 & 1 \nl
DDO  168 &   2.7 &   98 &  24.2 &  55 & 1 \nl
DDO  170 &   2.3 &   75 &  19.2 &  66 & 1 \nl
NGC   55 &   7.9 &  146 &   6.4 &  87 & 1 \nl
NGC  247 &   7.3 &  136 &   6.5 & 108 & 1 \nl
NGC  300 &   6.3 &  132 &   8.3 &  97 & 1 \nl
NGC  801 &  74.3 &  302 &   0.3 & 222 & 1 \nl
NGC 1003 &   9.7 &  133 &   3.5 & 115 & 1 \nl
NGC 1560 &   6.8 &  133 &   7.2 &  79 & 1 \nl
NGC 2403 &   6.6 &  154 &  10.2 & 136 & 1 \nl
NGC 2841 &  21.7 &  308 &   3.7 & 323 & 1 \nl
NGC 2903 &   3.2 &  166 &  51.0 & 201 & 1 \nl
NGC 2998 &  24.8 &  242 &   1.8 & 214 & 1 \nl
NGC 3109 &   8.7 &  141 &   4.9 &  67 & 1 \nl
NGC 3198 &   7.6 &  156 &   7.8 & 157 & 1 \nl
NGC 5033 &   5.9 &  170 &  15.2 & 222 & 1 \nl
NGC 5533 &  34.6 &  255 &   1.0 & 273 & 1 \nl
NGC 5585 &   1.8 &   99 &  56.9 &  92 & 1 \nl
NGC 6503 &   2.5 &  115 &  38.6 & 121 & 1 \nl
NGC 6674 & 119.5 &  655\tablenotemark{e} &   0.6 & 266 & 1 \nl
NGC 7331 & 103.0 &  982\tablenotemark{e} &   1.7 & 241 & 1 \nl
UGC 2259 &   6.1 &  137 &   9.5 &  90 & 1 \nl
UGC 2885 &  44.9 &  382\tablenotemark{e} &   1.3 & 298 & 1 \nl
NGC 3726 &   7.1 &  169 &  10.0 & 127 & 2 \nl
NGC 3877 &   4.8 &  171 &  23.0 & 139 & 2 \nl
NGC 3949 &   2.1 &  180 & 141.0 & 111 & 2 \nl
NGC 3953 &  10.1 &  228 &   9.0 & 205 & 2 \nl
NGC 3972 &   2.4 &  144 &  69.0 &  72 & 2 \nl
NGC 3992 &  10.6 &  235 &   9.0 & 251 & 2 \nl
NGC 4013 &   6.4 &  179 &  15.0 & 188 & 2 \nl
NGC 4085 &   2.1 &  172 & 121.0 &  65 & 2 \nl
NGC 4100 &   2.2 &  153 &  90.0 & 151 & 2 \nl
NGC 4138 &   1.3 &  135 & 193.0 & 174 & 2 \nl
NGC 4157 &   9.3 &  199 &   9.0 & 179 & 2 \nl
NGC 4217 &   2.2 &  164 &  99.0 & 138 & 2 \nl
UGC 6399 &   2.8 &   89 &  19.0 &  59 & 2 \nl
UGC 6466 &   1.3 &   74 &  64.0 &  40 & 2 \nl
UGC 6667 &   3.0 &   85 &  15.0 &  59 & 2 \nl
NGC 3917 &   3.6 &  124 &  22.0 & 104 & 2 \nl
UGC 6917 &   1.9 &  103 &  55.0 &  61 & 2 \nl
UGC 6923 &   1.8 &   96 &  55.0 &  36 & 2 \nl
NGC 4010 &   3.1 &  145 &  41.0 &  63 & 2 \nl
UGC 6983 &   8.4 &  119 &   4.0 &  94 & 2 \nl
UGC 7089 &   3.5 &   89 &  12.0 &  40 & 2 \nl
NGC 4183 &   5.3 &  105 &   7.0 &  90 & 2 \nl
F563-V2\tablenotemark{f}  &  0.94 &  118 & 283.0 & 110 & 3 \nl
F568-1\tablenotemark{f}   &   1.5 &  150 & 181.0 & 130 & 3 \nl
F568-3\tablenotemark{f}   &   2.5 &  129 &  48.0 & 100 & 3 \nl
F568-3   &   3.0 &  116 &  27.0 & 100 & 3 \nl
F568-V1\tablenotemark{f}  &   1.2 &  122 & 188.0 & 120 & 3 \nl
F568-V1  &   6.7 &  112 &   5.0 & 120 & 3 \nl
F574-1\tablenotemark{f}   &   1.5 &   86 &  92.0 &  90 & 3 \nl
F574-1   &   3.4 &   44 &   3.0 &  90 & 3 \nl
NGC 4123 &   6.3 &  101 &   4.7 & 130 & 4 \nl
NGC 5585 &   4.3 &   76 &  24.0 &  92 & 5 \nl
\enddata

\tablenotetext{a}{Fitted core radius in kpc}
\tablenotetext{b}{Fitted asymptotic velocity in km s$^{-1}$}
\tablenotetext{c}{Fitted central DM density in $10^{-3}$M$_\odot$ pc$^{-3}$}
\tablenotetext{d}{Observed maximum circular velocity in km s$^{-1}$}
\tablenotetext{e}{Adopted $V_{\rm flat}$ reduced to observed $V_{\rm max}$}
\tablenotetext{f}{``No disk'' fit}
\tablerefs{(1) Broeils 1992 p~244; (2) Verheijen 1997 p~246; (3) Swaters \etal\ 
2000; (4) Weiner \etal\  2000; (5) Blais-Ouellete \etal\ 1999}
\label{tab:data}
\end{deluxetable}

\end{document}